\documentclass{article}
\usepackage{amsmath, epsfig, citesort}
\usepackage{amssymb}
\usepackage{amsfonts}
\usepackage{latexsym}
\usepackage{longtable}
\usepackage{tabularx}

\newtheorem{theorem}{Theorem}

\newtheorem{definition}[theorem]{Definition}
\newtheorem{lemma}[theorem]{Lemma}

\newcommand{\qed}{{\hfill\rule{4pt}{7pt}}}
\def\pf{\noindent {\it Proof.} }

\title{Invertible binary matrix \\with maximum number of $2$-by-$2$ invertible submatrices}

\author{Yiwei Zhang$^{\text{a}}$, Tao Zhang$^{\text{a}}$, Xin Wang$^{\text{a}}$ and Gennian Ge$^{\text{b,c,}}$\thanks{Corresponding author (e-mail: gnge@zju.edu.cn). Research supported by the National Natural Science Foundation of China under Grant Nos. 61171198, 11431003 and 61571310, and the Importation and Development of High-Caliber Talents Project of Beijing Municipal Institutions.}\\
\footnotesize $^{\text{a}}$ School of Mathematical Sciences, Zhejiang University, Hangzhou 310027, Zhejiang, China\\
\footnotesize $^{\text{b}}$ School of Mathematical Sciences, Capital Normal University, Beijing 100048, China\\
\footnotesize $^{\text{c}}$ Beijing Center for Mathematics and Information Interdisciplinary Sciences, Beijing 100048, China.\\}

\begin{document}

\date{}\maketitle

\begin{abstract}
The problem is related to all-or-nothing transforms (AONT) suggested by Rivest as a preprocessing for encrypting data with a block cipher. Since then there have been various applications of AONTs in cryptography and security. D'Arco, Esfahani and Stinson posed the problem on the constructions of binary matrices for which the desired properties of an AONT hold with the maximum probability. That is, for given integers $t\le s$, what is the maximum number of $t$-by-$t$ invertible submatrices in a binary matrix of order $s$? For the case $t=2$, let $R_2(s)$ denote the maximal proportion of 2-by-2 invertible submatrices. D'Arco, Esfahani and Stinson conjectured that the limit is between 0.492 and 0.625. We completely solve the case $t=2$ by showing that $\lim_{s\rightarrow\infty}R_2(s)=0.5$.

\medskip


\smallskip


\smallskip

\end{abstract}

\section{Introduction}

The original motivation of this problem traces back to \cite{Rivest}, in which Rivest suggested using all-or-nothing transforms as a preprocessing for encrypting data with a block cipher, in the setting of computational security. Later Stinson \cite{Stinson} changed the setting to unconditional security and the generalized version of all-or-nothing transforms is defined in \cite{D'Arco} as follows:
\begin{definition}
Let $X$ be a finite set known as an alphabet. Let $s$ be a positive integer and consider a map $\phi: X^s \rightarrow X^s$. For an input $s$-tuple, say $x=(x_1,\dots,x_s)$, $\phi$ maps it to an output $s$-tuple, say $y=(y_1,\dots,y_s)$, where $x_i,y_i\in X$ for $1\le i \le s$. The map $\phi$ is an unconditionally secure {\it t-all-or-nothing transform} provided that the following properties are satisfied:

$\bullet$ $\phi$ is a bijection.

$\bullet$ If any $s-t$ out of the $s$ output values $y_1,\dots,y_s$ are fixed, then any $t$ of the input values $x_i$ $(1\le i \le s)$ are completely undetermined, in an information-theoretic sense.
\end{definition}

We will call such a function $\phi$ as a $(t,s,v)$-AONT, where $v=|X|$. And when $s$ and $v$ are clear or not relevant, we just call it a $t$-AONT.

What Rivest defined in \cite{Rivest} corresponds to the special case $t=1$. 1-AONT can provide a preprocessing called ``package transform" for block ciphers. Suppose we want to encrypt plaintexts $(x_1,\dots,x_s)$. Firstly we apply a 1-AONT to get $(y_1,\dots,y_s)=\phi(x_1,\dots,x_s)$. Note that the transform $\phi$ is not necessarily private. Then we encrypt $(y_1,\dots,y_s)$ using a block cipher and get the ciphertexts $z_i=e_K(y_i)$ for $1\le i \le s$, where $e_K$ is the encryption function. The receiver can decrypt the ciphertexts and then use the inverse transform $\phi^{-1}$ to restore the plaintexts. However, any adversary needs to decrypt the whole ciphertexts and get the exactly values of $(y_1,\dots,y_s)$ (by means of exhaustive key search, say) in order to determine any one symbol among the plaintexts. In other words, a partial decryption cannot provide any information about each symbol among the plaintexts due to the property of a 1-AONT. In this sense, the application of 1-AONTs gives a certain amount of additional security over block ciphers. Extensions of this technique are studied in \cite{Canda,Desai}. AONTs also have various other applications in cryptography and security. For example, it is applied in network coding \cite{Cascella,Guo}, secure data transfer \cite{Vasudevan}, anti-jamming techniques \cite{Proano}, exposure-resilient functions \cite{Canetti}, secure distributed cloud storage \cite{Liu,Song}, secure secret sharing schemes \cite{Resch} and query anonymization for location-based services \cite{Zhang}.

However, the properties of 1-AONT do not say anything regarding the partial information that might be revealed about more than one of the $s$ input values. Say, it is possible to derive the sum of two input values with only some relatively small number of output values. That is exactly the motivation for the general definition of a $t$-AONT. Similarly as above, if a $t$-AONT is applied before using a block cipher, then the adversary will have no information regarding any boolean function of any $t$ symbols among the plaintexts, unless he could do enough decryption to get more than $s-t$ symbols in $(y_1,\dots,y_s)$.

Linear AONTs are of particular interests. Let the alphabet set be $\mathbb{F}_q$, the finite field of order $q$. A $(t,s,q)$-AONT with alphabet $\mathbb{F}_q$ is {\it linear} if each $y_i$ is an $\mathbb{F}_q$-linear function of $(x_1,\dots,x_n)$. Then it can be represented by an invertible $s$-by-$s$ matrix with entries from $\mathbb{F}_q$. In \cite{D'Arco} it was shown that the second property of AONT requires that every $t$-by-$t$ submatrices of $M$ must also be invertible. Matrices satisfying this condition do exist when $q$ is a prime power and $q\ge 2s$ \cite{D'Arco}. However, when we are working on $\mathbb{F}_2$, it is easy to see that there is no linear $(1,s,2)$-AONT for $s>1$ and no linear $(2,s,2)$-AONT for $s>2$. So in \cite{D'Arco} D'Arco et al. proposed the question on how close one can get to an AONT. That is, for given integers $t\le s$, what is the maximum number of $t$-by-$t$ invertible submatrices in a binary matrix $M$ of order $s$? In the sequel, invertibility of a matrix refers to the invertibility over $\mathbb{F}_2$. The following notations are defined in \cite{D'Arco}:
\begin{equation*}
N_t(M)= \text{ number of invertible } t \text{-by-} t \text{ submatrices of } M,
\end{equation*}
\begin{equation*}
R_t(M)=\frac{N_t(M)}{{s \choose t}^2},
\end{equation*}
\begin{equation*}
R_t(s)=\max \{R_t(M): M \text{ is an } s \text{-by-} s \text{ invertible binary matrix}\}.
\end{equation*}

$R_1(s)=1-\frac{s-1}{s^2}$ was shown in \cite{D'Arco}. Also in \cite{D'Arco} upper and lower bounds for $R_2(s)$ are analyzed and $\lim_{s\rightarrow\infty}R_2(s)$ is conjectured to exist and the value is between 0.494 and 0.625. In this paper, we give a complete solution by showing that $\lim_{s\rightarrow\infty}R_2(s)=0.5$. The rest of the paper is organized as follows. In Section \ref{upperbound} we analyze the upper bound of $R_2(s)$ and show that $\lim_{s\rightarrow\infty}R_2(s)\le0.5$. In Section \ref{lowerbound} we use some probabilistic tools to derive the lower bound for $R_2(s)$ and show that $\lim_{s\rightarrow\infty}R_2(s)\ge0.5$. In Section \ref{construction} we offer a construction via cyclotomy as an illustrative example. Finally we conclude in Section \ref{conclude}.

\section{Upper bound via integer programming} \label{upperbound}
In this section we analyze the upper bound of $R_2(s)$. The restriction that the matrix itself should be invertible can be ignored for the moment. Note that a 2-by-2 binary matrix is invertible if and only if it is one of the following matrices:
\begin{equation*}
\left(
  \begin{array}{cc}
    1 & 0\\
    0 & 1\\
  \end{array}
\right),
\left(
  \begin{array}{cc}
    0 & 1\\
    1 & 0\\
  \end{array}
\right),
\left(
  \begin{array}{cc}
    1 & 1\\
    1 & 0\\
  \end{array}
\right),
\left(
  \begin{array}{cc}
    1 & 1\\
    0 & 1\\
  \end{array}
\right),
\left(
  \begin{array}{cc}
    1 & 0\\
    1 & 1\\
  \end{array}
\right),
\left(
  \begin{array}{cc}
    0 & 1\\
    1 & 1\\
  \end{array}
\right).
\end{equation*}

Firstly we focus on those matrices containing exactly $c$ entries as ``1", for a given integer $c$. We now analyze how these ``1" entries should be distributed in order to get the maximal number of 2-by-2 invertible submatrices. For $1\le i \le n$, let $x_i$ be the weight of each row and let $y_i$ be the weight of each column. For $1\le i <j \le n$, let $z_{i,j}$ be the intersection number of the $i$-th and $j$-th row. That is, $z_{i,j}=|\{k:M_{i,k}=M_{j,k}=1\}|$. Besides the natural restriction $\sum_{i=1}^n x_i=\sum_{i=1}^n y_i=c$, a standard double counting argument also implies a furthur restriction that $\sum_{1\le i<j\le n} z_{i,j}=\sum_{i=i}^n {y_i \choose 2}$. Then the number of invertible 2-by-2 submatrices provided by the $i$-th and $j$-th row can be easily calculated as
\begin{equation*}
z_{i,j}(x_i-z_{i,j})+z_{i,j}(x_j-z_{i,j})+(x_i-z_{i,j})(x_j-z_{i,j})=x_ix_j-z_{i,j}^2.
\end{equation*}

So we are actually facing an integer programming problem as follows.

\begin{align*}
\text{maximize}: &\sum_{1\le i < j \le s} x_ix_j-z_{i,j}^2\\
\text{subject to}:  &\sum_{1\le i \le s} x_i=\sum_{1\le i \le s} y_i=c,\\
& \sum_{1\le i <j\le s} z_{i,j}=\sum_{1\le i \le s} {y_i \choose 2},\\
& x_i,y_i,z_{i,j} \in \mathbb{N}, 1\le i<j\le s, \\
& c \in \mathbb{N}, 0 \le c \le s^2.
\end{align*}

The solution to the integer programming problem, divided by the factor ${s \choose 2}^2$, serves as an upper bound of $R_2(s)$. Whether or not the exact value can be reached is not a trivial problem since an invertible matrix with the exact parameters $x_i,y_i,z_{i,j}$ may not exist. Generally the exact maximum value of $R_2(s)$ for each $s$ should follow a case-by-case analysis and we will show for $s=10$ as an illustrative example. We denote a multiset in the form of $a_1^{n_1}\dots a_k^{n_k}$, indicating that the element $a_i$ appears $n_i$ times in the multiset. For the case $s=10$, the integer programming suggests the maximum value 1216 might be achieved when the row weights and column weights are multisets of the form $8^27^8$ and the set of intersection numbers is of the form $5^{44}4^1$. However, this certainly could not happen since the two rows of weight 8 will certainly have their intersection number at least 6. We take one step back to search for an invertible matrix with 1215 2-by-2 invertible submatrices, with row weights and column weights being $7^{10}$ and the set of intersection numbers being $5^{30}4^{15}$. By setting $S=\{1,2,4\}$, the binary matrix $M$ with $M_{i,j}=1$ if and only if $i-j\notin S \pmod{10}$ is the desired matrix as follows. So we have $R_2(10)=\frac{1215}{2025}=0.6$.
$$\left(
  \begin{array}{cccccccccc}
1&1&1&1&1&1&0&1&0&0\\
0&1&1&1&1&1&1&0&1&0\\
0&0&1&1&1&1&1&1&0&1\\
1&0&0&1&1&1&1&1&1&0\\
0&1&0&0&1&1&1&1&1&1\\
1&0&1&0&0&1&1&1&1&1\\
1&1&0&1&0&0&1&1&1&1\\
1&1&1&0&1&0&0&1&1&1\\
1&1&1&1&0&1&0&0&1&1\\
1&1&1&1&1&0&1&0&0&1\\
  \end{array}
\right)$$

At this moment, the asymptotic solution will be enough for deriving the upper bounds for $\lim_{s\rightarrow\infty}R_2(s)$. Consider the slack version of the integer programming above. It can be easily seen that the maximum value is achieved when each set $\{x_i\}_{1\le i \le s}$, $\{y_i\}_{1\le i \le s}$ and $\{z_{i,j}\}_{1\le i <j\le s}$ is averagely distributed. That is, for a given $c$, $x_i=y_i=\frac{c}{s}$, and $z_{i,j}=\frac{c(c-s)}{s^2(s-1)}$ and the objective function attains ${s\choose 2}(\frac{c^2}{s^2}-\frac{c^2(c-s)^2}{s^4(s-1)^2})$. This value achieves the maximum at $c=\frac{3s+\sqrt{8s^4-16s^3+9s^2}}{4}$. So asymptotically set $c\sim \frac{\sqrt{2}s^2}{2}$ and we finally derive:

\begin{theorem} \label{up}
$\lim_{s\rightarrow\infty}R_2(s)\le0.5$.
\end{theorem}

\section{Lower bound via probabilistic method} \label{lowerbound}
It has been suggested in \cite{D'Arco} that random construction may lead to very good results. Selecting each entry independently to be ``1" with probability $p=\sqrt{\frac{1}{2}}$ will result in a matrix with the expectation of its 2-by-2 submatrices to be $\frac{1}{2}{s\choose 2}{s\choose 2}$. The only problem is to decide whether the matrix itself is invertible or not. In fact, the invertibility of the whole matrix is not a severe problem. We offer two ways to deal with this potential flaw. One will be illustrated in the next section, and the other one below is a combination of a standard deviation argument and a powerful result obtained in \cite{Cooper}.

\begin{lemma}{\rm\cite{Cooper}} \label{constant}
Let $M$ be a random $s$-by-$s$ matrix in $\mathbb{F}_2$. Each entry is independently and identically distributed, with $\Pr[M_{i,j}=1]=p(s)$ and $\Pr[M_{i,j}=0]=1-p(s)$. If $\min \{{p(s),1-p(s)}\} \ge (\log{s}+d(s))/s$ for any $d(s)\rightarrow \infty$, then $\Pr[M \text{ is invertible}]$ tends to a constant $c\approx 0.28879$.
\end{lemma}

\begin{theorem} \label{low}
Given arbitrary positive $\epsilon>0$, there exists a sufficiently large $S$. For every $s>S$, there exists an invertible binary matrix $M$ with $R_2(M)>\frac{1}{2}-\epsilon$. Then, $\lim_{s\rightarrow\infty}R_2(s)\ge 0.5$.
\end{theorem}
\pf
For $1\le i <j\le s$, $1\le k<l \le s$, let $X_{i,j;k,l}$ be the indicating random variable of the event ``the 2-by-2 submatrix induced by the $i$-th and $j$-th row with the $k$-th and $l$-th column is invertible". That is, $X_{i,j;k,l}=1$ if the event occurs and otherwise $X_{i,j;k,l}=0$. Let $X=\sum_{1\le i<j\le s}\sum_{1\le k<l \le s}X_{i,j;k,l}$. $X$ counts the total number of invertible 2-by-2 submatrices. Let each entry of the matrix be independently and identically distributed, each chosen to be ``1" with probability $p$. By the linearity of expectation, we have
\begin{equation*}
E[X]=\sum_{1\le i<j\le s}\sum_{1\le k<l \le s} E[X_{i,j;k,l}]= {s \choose 2}^2 (4p^3(1-p)+2p^2(1-p)^2)= {s \choose 2}^2 (2p^2-2p^4).
\end{equation*}
By choosing $p=\sqrt{\frac{1}{2}}$, the expectation is maximized as $E[X]=\frac{1}{2}{s \choose 2}^2$.
Now we consider the variance of the random variable $X$. Notice that the covariance between two indicating random variables $X_1:=X_{i_1,j_1;k_1,l_1}$ and $X_2:=X_{i_2,j_2;k_2,l_2}$ can be calculated as
\begin{equation*}
\text{Cov}[X_1,X_2]=E[(X_1-E[X_1])(X_2-E[X_2])]=\frac{1}{4}(\Pr[X_1=X_2]-\Pr[X_1\neq X_2])=\frac{1}{4}(2\Pr[X_1=X_2]-1).
\end{equation*}
There are three different situations according to the intersections of the two submatrices:

1) The two submatrices are disjoint. Then they are independent and the covariance is zero.

2) The two submatrices intersect at only one entry. The calculation can be further divided into two parts:

$\bullet$ Let $\mathcal{A}$ be the event that the common entry is 0 with probability $1-p$. Under this condition,
\begin{align*}
&\Pr[X_1=1|\mathcal{A}]=\Pr[X_2=1|\mathcal{A}]=p^2(1-p)+p^3=\frac{1}{2},\\
&\Pr[X_1=X_2|\mathcal{A}]=\Pr[X_1=1|\mathcal{A}]\Pr[X_2=1|\mathcal{A}]+\Pr[X_1=0|\mathcal{A}]\Pr[X_2=0|\mathcal{A}]=\frac{1}{2}\times\frac{1}{2}+\frac{1}{2}\times\frac{1}{2}=\frac{1}{2}. \\
\end{align*}

$\bullet$ Let $\mathcal{B}$ be the event that the common entry is 1 with probability $p$. Under this condition,
\begin{align*}
&\Pr[X_1=1|\mathcal{B}]=\Pr[X_2=1|\mathcal{B}]=3p^2(1-p)+p(1-p)^2=\frac{1}{2},\\
&\Pr[X_1=X_2|\mathcal{B}]=\Pr[X_1=1|\mathcal{B}]\Pr[X_2=1|\mathcal{B}]+\Pr[X_1=0|\mathcal{B}]\Pr[X_2=0|\mathcal{B}]=\frac{1}{2}\times\frac{1}{2}+\frac{1}{2}\times\frac{1}{2}=\frac{1}{2}. \\
\end{align*}

Then the covariance of this kind of unordered pair is also zero.

3) The two submatrices intersect at two entries (the two common entries are either in a same row or in a same column). Note that for each $X_{i,j;k,l}$, the number of other indicating variables that intersect with it at two entries is $4(s-2)$. So there are totally $2(s-2){s\choose 2}^2$ such unordered pairs. The calculation can be further divided into three parts:

$\bullet$ Let $\mathcal{C}$ be the event that the two common entries are both 1 (this occurs with probability $p^2$). Under this condition,
\begin{align*}
&\Pr[X_1=1|\mathcal{C}]=\Pr[X_2=1|\mathcal{C}]=2p(1-p)=2p-1,\\
&\Pr[X_1=X_2|\mathcal{C}]=\Pr[X_1=1|\mathcal{C}]\Pr[X_2=1|\mathcal{C}]+\Pr[X_1=0|\mathcal{C}]\Pr[X_2=0|\mathcal{C}]=(2p-1)^2+(2-2p)^2=9-12p. \\
\end{align*}

$\bullet$ Two common entries are both 0 (this occurs with probability $(1-p)^2$), then both $X_1$ and $X_2$ are always zero and $\Pr[X_1=X_2]=1$;

$\bullet$ Let $\mathcal{D}$ be the event that exactly one of the common entries is 1 (this occurs with probability $2p(1-p)$). Under this condition,
\begin{align*}
&\Pr[X_1=1|\mathcal{D}]=\Pr[X_2=1|\mathcal{D}]=p^2+pq=p,\\
&\Pr[X_1=X_2|\mathcal{D}]=\Pr[X_1=1|\mathcal{D}]\Pr[X_2=1|\mathcal{D}]+\Pr[X_1=0|\mathcal{D}]\Pr[X_2=0|\mathcal{D}]=p^2+(1-p)^2=2-2p. \\
\end{align*}

Then the covariance of this kind of unordered pair is
\begin{equation*}
\frac{1}{4}(p^2\times(17-24p)+(1-p)^2\times1+2p(1-p)\times(3-4p))=\frac{3}{4}-p.
\end{equation*}

Now we can proceed to calculate the variance of $X$.
\begin{equation*}
\text{Var}[X]=\sum_{1\le i<j\le s}\sum_{1\le k<l \le s} \text{Var}[X_{i,j;k,l}] + 2(s-2){s\choose 2}^2\times(\frac{3}{4}-p)\sim s^5.
\end{equation*}

We now apply the Chebyshev's Inequality. For any positive $\lambda$, $\Pr[|X-E[X]|\ge\lambda\sigma]\le\frac{1}{\lambda^2}$, where $\sigma=\sqrt{\text{Var}[X]}$ is the standard deviation. For arbitrary positive $\epsilon>0$, set $\lambda\sigma=\epsilon{s\choose 2}^2$. Since $\sigma\sim s^{5/2}$, so $\lambda\sim s^{3/2}$. Then $\Pr[|X-E[X]|\ge \epsilon{s\choose 2}^2]\le \frac{1}{s^3}\rightarrow 0~(s\rightarrow\infty)$. Then for sufficiently large $s$, by Lemma \ref{constant}, $\Pr[M\text{ is invertible}]+\Pr[X>(\frac{1}{2}-\epsilon){s\choose 2}^2]>0$. So $R_2(s)>\frac{1}{2}-\epsilon$ for arbitrary $\epsilon>0$ and then $\lim_{s\rightarrow\infty}R_2(s)\ge 0.5$.\qed

Finally Theorem \ref{up} and Theorem \ref{low} together lead to our main result:
\begin{theorem}
Let $R_2(s)$ be the maximal proportion of $2$-by-$2$ invertible submatrices in an $s$-by-$s$ invertible binary matrix. We have
\begin{equation*}
\lim_{s\rightarrow\infty}R_2(s)=0.5.
\end{equation*}
\end{theorem}

\section{Explicit constructions} \label{construction}

The probabilistic result does not indicate how to construct such a matrix and a direct derandomization approach seems to be unrealistic since the conditional expectation when selecting each entry is hard to compute. In this section, we suggest a direct approach to construct a class of near-optimal matrices.

The construction consists of a main step in which the essential frame of the matrix $M$ is formed and then an adjusting step which guarantees the invertibility of $M$ while simultaneously does not affect the value of $R_2(M)$ asymptotically.

Recall the proof considering the upper bound, we see that an optimal matrix should have balanced row weights, balanced column weights and also balanced intersection numbers between rows. Recall the proof considering the lower bound, we see that the total number of ``1" entries should be close to $\sqrt{\frac{1}{2}}s^2$. The seek for optimal matrices should follow these two instructions.

Let $\mathcal{S}$ be a subset of $\{1,2,\dots,s\}$. We construct the matrix $M$ by setting $M_{i,j}=1$ if and only if $i-j\notin\mathcal{S},\pmod{s}$. The two instructions above indicate that:

$\bullet$ The size of $\mathcal{S}$ is about $(1-\sqrt{\frac{1}{2}})s$.

$\bullet$ Let $\Delta(\mathcal{S})$ be the multiset $\{x-y:x,y\in \mathcal{S},x\neq y\}$. Let $m_i$ be the multiplicity of $i$ in $\Delta(\mathcal{S})$, $1\le i \le s-1$. These multiplicities should be almost equal.

The key is to find an appropriate $\mathcal{S}$, where cyclotomy turns out to be very useful. Actually, in \cite{D'Arco} the authors used cyclotomic classes of order 4 to construct a class of matrices with $R_2(M)\approx 0.492$. We mimic their ideas, using cyclotomic classes of order 7, since the size of the corresponding $\mathcal{S}$ will be closer to the desired size.

\subsection{Main step: constructions by cyclotomy}

Let $p$ be a prime and $\gamma$ be a fixed primitive element of $\mathbb{F}_p$. Let $N>1$ be a divisor of $p-1$. We define the $N$th cyclotomic classes $C_{0},C_{1},\ldots,C_{N-1}$ of $\mathbb{F}_p$ by
$$C_{i}=\left\{\gamma^{jN+i}|0\leq j\leq\frac{p-1}{N}-1\right\},$$
where $0\leq i\leq N-1$. That is, $C_{0}$ is the $N$-th power residues modulo $p$, and $C_{i}=\gamma^{i}C_{0}$, $1\leq i\leq N-1$. For integers $i,j$ with $0\leq i,j<N$, the cyclotomic number of order $N$ is defined by
$$(i,j)_{N}=|(C_{i}+1)\bigcap  C_{j}|.$$

The following lemma summarizes some basic properties of cyclotomic numbers.
\begin{lemma}{\rm \cite{Berndt}}\label{lemcyc}
Let $p=ef+1$ be some odd prime. Then
\begin{enumerate}
  \item $(i,j)_{e}=(i',j')_{e}$, when $i\equiv i'\pmod{e}$ and $j\equiv j'\pmod{e}$.
  \item $ (i,j)_{e}
    =\begin{cases}(j,i)_{e}, & \textup{ if } f \textup{ even};\\
    (j+e/2,i+e/2)_{e}, & \textup{ if } f \textup{ odd}.\end{cases}$
  \item $\sum_{i=0}^{e-1}(i,j)_{e}=f-\delta_{j}$, where $\delta_{j}=1$ if $j\equiv0\pmod{e}$; otherwise $\delta_{j}=0$.
\end{enumerate}
\end{lemma}

Within this section, we always assume that $p=7f+1$ is a prime. We need the following result about cyclotomic numbers of order $7$.
\begin{lemma}{\rm \cite{Leonard,Leonard1}}\label{thmcyc}
If $p\equiv1\pmod{7}$, then for $0\leq i,j\leq6$, $\lim_{p\rightarrow\infty}\frac{(i,j)_{7}}{p}=\frac{1}{49}$.
\end{lemma}

Now we set the matrix $M'=(m_{ij})$ to be the $p\times p$ matrix on $\mathbb{F}_2$ with rows and columns indexed by $\mathbb{F}_p$, and
$$m_{ij}=\begin{cases}1, & \textup{ if } j-i\in C_{0}\bigcup C_{1};\\
    0, & \textup{ otherwise}.\end{cases}$$
Let $M$ be the complement of $M'$.

We first consider the matrix $M'$. It is obvious that the number of invertible 2-by-2 submatrices contained in rows $i_{1}$ and $i_{2}$ only depends on the number $i_{1}-i_{2}$. Then we consider rows $0$ and $i$ of $M'$. Define
$$n_{i}=\{j|m_{0j}=m_{ij}=1\}.$$
Then we can compute that
 \begin{align*}
n_{i}&=|(C_{0}\bigcup C_{1})\bigcap((C_{0}\bigcup C_{1})+i)|\\
     &=|C_{0}\bigcap(C_{0}+i)|+|C_{0}\bigcap(C_{1}+i)|+|C_{1}\bigcap(C_{0}+i)|+|C_{1}\bigcap(C_{1}+i)|\\
     &=|i^{-1}C_{0}\bigcap(i^{-1}C_{0}+1)|+|i^{-1}C_{0}\bigcap(i^{-1}C_{1}+1)|+|i^{-1}C_{1}\bigcap(i^{-1}C_{0}+1)|+|i^{-1}C_{1}\bigcap(i^{-1}C_{1}+1)|.
\end{align*}
Let $i^{-1}C_{0}=C_{m}$, for some $0\leq m\leq6$, then $i^{-1}C_{1}=C_{m+1}$. By Lemma~\ref{lemcyc}, we have
 \begin{align*}
n_{i}&=|C_{m}\bigcap(C_{m}+1)|+|C_{m}\bigcap(C_{m+1}+1)|+|C_{m+1}\bigcap(C_{m}+1)|+|C_{m+1}\bigcap(C_{m+1}+1)|\\
     &=(m,m)_{7}+(m,m+1)_{7}+(m+1,m)_{7}+(m+1,m+1)_{7},\\
     &=(m,m)_{7}+2(m,m+1)_{7}+(m+1,m+1)_{7}.
\end{align*}

In rows $0$ and $i$ of $M$, suppose there are $a_{0}$ occurrences of $\left(
                                                                       \begin{array}{c}
                                                                         0 \\
                                                                         0 \\
                                                                       \end{array}
                                                                     \right)
$,  $a_{1}$ occurrences of $\left(
                                                                       \begin{array}{c}
                                                                         0 \\
                                                                         1 \\
                                                                       \end{array}
                                                                     \right)
$,  $a_{2}$ occurrences of $\left(
                                                                       \begin{array}{c}
                                                                         1 \\
                                                                         0 \\
                                                                       \end{array}
                                                                     \right)
$ and  $a_{3}$ occurrences of $\left(
                                                                       \begin{array}{c}
                                                                         1 \\
                                                                         1 \\
                                                                       \end{array}
                                                                     \right)
$. Note that $M$ is the complement of $M'$, we have
 \begin{align*}
 &a_{0}=n_{j},\\
 &a_{1}=a_{2}=2f-n_{j},\\
 &a_{3}=p-4f+n_{d}=3f+1+n_{j}.
\end{align*}
Thus we obtain
$$a_{1}a_{2}+a_{1}a_{3}+a_{2}a_{3}=16f^{2}-6fn_{j}+4f-n_{j}^{2}-2n_{j}.$$
Therefore the total number of invertible 2-by-2 submatrices in $M$ is
 \begin{align*}
&\frac{p(p-1)}{14}\sum_{m=0}^{6}(16f^{2}-6fn_{j}+4f-n_{j}^{2}-2n_{j})\\
&=\frac{p(p-1)}{14}\sum_{m=0}^{6}(16f^{2}+4f-(6f+2)((m,m)_{7}+2(m,m+1)_{7}+(m+1,m+1)_{7})-\\
&((m,m)_{7}+2(m,m+1)_{7}+(m+1,m+1)_{7})^{2}).
\end{align*}
From Lemma~\ref{thmcyc}, we see that, for all $0\leq m\leq6$, $\frac{(m,m)_{7}}{f}=\frac{(m,m+1)_{7}}{f}=\frac{1}{7}$ if $f$ approaches infinity, then we have
 \begin{align*}
&\lim_{p\rightarrow\infty}\frac{2}{7p(p-1)}\sum_{m=0}^{6}(16f^{2}+4f-(6f+2)((m,m)_{7}+2(m,m+1)_{7}+(m+1,m+1)_{7})-\\
&((m,m)_{7}+2(m,m+1)_{7}+(m+1,m+1)_{7})^{2})\\
&=\lim_{f\rightarrow\infty}\frac{2}{7(7f+1)7f}\sum_{m=0}^{6}(16f^{2}+4f-(6f+2)(\frac{4f}{7}+o(f))-(\frac{4f}{7}+o(f))^{2})\\
&=\frac{1200}{2401}.
\end{align*}

\subsection{Adjusting step}

It is still necessary to decide whether the construction above gives an invertible matrix. We claim that even if the matrix $M$ is singular, we can make a few changes to turn it into an invertible matrix without affecting the value of $R_2(M)$ asymptotically.

\begin{lemma}
Any binary matrix $M$ can be adjusted to an invertible matrix via adjusting the entries on the diagonal.
\end{lemma}

\pf
We prove by induction. Firstly, set the entry $M_{1,1}$ to be ``1".
If we have set the leading principal minor of order $k$ (denoted as $P_k$) to be invertible, then now we look at the choice of $M_{k+1,k+1}$. To calculate the determinant of $P_{k+1}$, we expand $M_{k+1,k+1}$ along the $(k+1)$-th row. The summation contains a part $\det(P_k)\cdot M_{k+1,k+1}$. Since by induction $\det(P_k)\neq0$, then we can adjust the entry $M_{k+1,k+1}$ to guarantee that $\det(P_{k+1})\neq0$.
\qed

Only at most $p$ entries are modified, each entry is contained in $(p-1)(p-1)$ 2-by-2 submatrices, so the number of 2-by-2 submatrices affected are no more than $p(p-1)^2\thicksim p^3=o(p^4)$. So for sufficiently large $p$, this amount can be neglected.

In summary, the main step via cyclotomy plus the adjustment on diagonals together lead to:
\begin{theorem}
For every prime $p$, $p\equiv 1\pmod{7}$, the construction above gives a matrix $M_p$, with
\begin{equation*}
\lim_{p\rightarrow\infty}R_2(M_p)= \frac{1200}{2401}\approx0.4997917.
\end{equation*}
\end{theorem}

\section{Conclusions} \label{conclude}

In this paper we show that $\lim_{s\rightarrow\infty}R_2(s)=0.5$ as a complete solution to the problem posed by D'Arco et al. \cite{D'Arco}. For the cases with larger $t$, the randomized construction plus the deviation argument still works for a lower bound of $\lim_{s\rightarrow\infty}R_t(s)$. For example, $\lim_{s\rightarrow\infty}R_3(s)\ge 0.38817$ (each entry is chosen as $1$ with probability $p\approx0.63056$). However, the similar idea regarding the upper bound via integer programming seems hard to analyze. We conjecture that for any $t\le s$, the exact value of $\lim_{s\rightarrow\infty}R_t(s)$ is exactly the lower bound derived from the probabilistic analysis, and the corresponding optimal matrix should be in a balanced form. That is, the intersection of every $r\le t$ rows (columns) is about $sp^r$, where $p$ is the optimal probability for choosing each entry as ``1".

\end{document}